\documentclass{article}

\begin{document}



\title{What did we learn from studying \\acoustic black holes ?}

\author{
Renaud Parentani\footnote{
Laboratoire de Math\'ematique et Physique Th\'eorique, CNRS-UMR 6083, 
Universit\'e de Tours, Parc de Grammont,
37200 Tours, France.}}


\maketitle                            

\begin{abstract}
The study of acoustic black holes has been undertaken to provide new 
insights about the role of high frequencies in black hole evaporation.
Because of the infinite gravitational redshift from the event horizon,
Hawking quanta emerge from configurations which possessed ultra high 
(trans-Planckian) frequencies. Therefore Hawking radiation cannot be 
derived within the framework of a low energy effective theory; 
and in all derivations there are some assumptions concerning Planck scale
physics. The analogy with condensed matter physics was thus introduced
to see if the asymptotic properties of the Hawking phonons emitted by 
an acoustic black hole, namely stationarity and thermality, 
are sensitive to the high frequency physics 
which stems from the granular character of matter
and which is governed by a non-linear dispersion relation. 
In 1995 Unruh showed that they are not sensitive in this respect, 
in spite of the fact that phonon propagation
near the (acoustic) horizon drastically differs from that of photons. 
In 2000 the same analogy was used to establish the robustness of 
the spectrum of primordial density fluctuations in inflationary models. 
This analogy is currently stimulating research for experimenting 
Hawking radiation. Finally it could also be a useful guide for 
going beyond the semi-classical description of black hole evaporation.


\end{abstract}


\subsection*{The motivation for the analogy}

The semi-classical description of black hole evaporation
is based on the hypothesis that
the dynamics of the system (black hole + radiation) 
can be well approximated by treating the gravitational field classically. 
In this case the quantized radiation field $\phi$
obeys the d'Alembert equation 
\begin{equation}
{1 \over \sqrt{-g} }\partial_\mu   \sqrt{-g}g^{\mu \nu }\partial_\nu
 \phi = 0\, ,
\end{equation}
where $g^{\mu \nu }$ is the metric of the collapsing body.  
Following Hawking\cite{hawk},
one obtains that an isolated black hole formed by gravitational
collapse radiates away its mass by emitting a thermal flux 
with a temperature given by $T= \kappa / 2\pi$ where $\kappa$ is 
its surface gravity.

The hypothesis of having treated $g^{\mu \nu }$
classically finds support in the fact that $\langle T_{\mu \nu } \rangle$,
the expectation value of the energy-momentum tensor of $\phi$,
is regular. 
By solving Einstein equations driven by $\langle T_{\mu \nu } \rangle$,
one then finds 
that the evaporation stays adiabatic as long as the residual mass 
of the hole is much larger than the Planck mass, see e.g. \cite{GO}. 

However, the validity of the semi-classical description becomes dubious when 
inquiring about the origin of Hawking quanta.
Eq. (1) predicts that they emerge from configurations 
localized extremely close to the horizon and characterized by 
ultra-high frequencies (when measured by an infalling 
observer). This is specific and
intrinsic to horizon quantum physics 
as it follows from free propagation near an event horizon 
and from second quantization. 
The infinite gravitational redshift from the event horizon 
implies that during the last stage of the 
collapse the radiation emitted by the body is redshifted according to
\begin{equation}
\lambda(t) \propto \Omega \ e^{-\kappa t} \, ,
\end{equation}
where $\Omega$ is the emitted frequency 
and $\lambda$ that received on ${\cal J}^+$.
Classically, there is thus no appeal to high frequencies
since the initial frequencies $\Omega$ are bounded.
On the contrary, Hawking quanta emerge from vacuum fluctuations 
with very high $\Omega$
precisely because these are redshifted as in Eq. (2). 
Indeed, the steady character of Hawking radiation 
is achieved at time $t$ by referring to vacuum frequencies which grow 
according to $\Omega(t) \propto \kappa e^{\kappa t}$,
see Eq. (3.34) in \cite{GO}. 

Hawking radiation therefore cannot be derived within the framework of a 
low energy effective theory. Moreover since Planck scale physics
is unknown,
it is important to determine if/why the asymptotic properties 
of Hawking radiation could be insensitive 
to it. From the outset\cite{unruh81}
this question has been the principle 
motivation for studying acoustic black holes.


\subsection*{The acoustic black hole geometry}

The concept of acoustic black hole arises from 
the analogy between light propagation in a curved background 
and sound propagation in a moving fluid\cite{unruh81,visser}.
Consider a vorticity-free flow of a barotropic fluid. The 
velocity field can be written as 
$v= \nabla \Psi$ and the pressure 
$P$ is related to the density $\rho$ by an 
equation of state $P=P(\rho)$. 
Consider also the equation of continuity and Euler equation. 
Then linearize these three equations about a background solution, 
 $v_B= \nabla \Psi_B $,  $P_B $,  $\rho_B$, 
and consider the equation for 
$\psi=\Psi- \Psi_B $, see \cite{visser} for details.
Verify that $\psi$ 
obeys Eq. (1) 
when introducing the $4$D acoustic metric $g_{\mu \nu}^{ac}$ 
 given (up to a conformal factor) 
by $ g_{tt}^{ac}=-(c^2_s -v_B^2), \, 
g_{ti}= -v_{B\, i}, \, g_{ij}= \delta_{ij}$. The velocity of sound $c_s$
is given by $c_s^2 = dP/d\rho$.


To obtain an acoustic geometry which behaves like that of
a black hole, one must consider a flow which becomes supersonic 
without developing any shock wave. A well-known device 
(by hydrodynamicians) which realizes this is the Laval nozzle:
 a pipe wherein $\vert v_B \vert$ reaches $c_s$ at its narrowest radius. 
When considering
a flow to the left in the $x$-direction, the velocity behaves as 
$v_B(x)= - c_s + x \kappa_{ac}$ and the acoustic horizon is located at $x=0$. 
In the near horizon region, for $x \ll c_s/\kappa_{ac}$, 
 null (retarded) geodesics behave like 
\begin{equation}
x(t)=x_0 \, e^{\kappa_{ac}(t-t_0)}\,\, .
\label{neweq}
\end{equation}
Hence, for $x < 0$, right moving sound waves 
are carried away to the left by the supersonic flow.
For $x > 0$ instead, they propagate to the right and 
their frequencies are redshifted
according to Eq. (2) with $\kappa=\kappa_{ac}$.

Thus, if the phonon field $\psi$ is second quantized and 
if its state is regular across the horizon, the standard
treatment\cite{hawk,GO} applies. It tells us that a constant flux
of phonons at temperature $T_{ac}= \kappa_{ac}/2 \pi$
will be emitted against the flow, to the right. 
The quantization of the phonon field is physically pertinent in ${\rm He}^4$
and in Bose-Einstein condensates (BEC). The regularity of the state
of $\psi$
requires that the flow $v_B(x)$ be regular across the horizon. 

\subsection*{The usefulness of the analogy}

So far we have not learned much because 
we have not yet taken into account the fact
that 
the dispersion relation of phonons 
is not linear when the wave length approaches the inter-atomic 
distance, $d_c$. 
Three possibilities can be obtained. Two can be modeled by 
\begin{equation}
\Omega^2  = c_s^2 k^2 [ 1 + \xi \, k^2 d_c^2 + O(k^4)] \, ,
\end{equation}
 with $ \xi = \pm 1$. $ \xi = 1$ is 
realized in BEC and leads to supersonic propagation. $ \xi = - 1$
is
realized  in ${\rm He}^4$ and leads to subsonic propagation.
The third case describes propagation in an absorptive medium
 such as a viscous fluid\cite{visser}. It is governed by a correction term,
the bracket in Eq. (4), given by $[1-i \vert k \vert d_c]$.

When taking into account the non-linearities of Eq. (4)
while deriving the production of Hawking phonons, one can {\it test} if
the asymptotic properties of the flux
are sensitive to these non-linearities.
In 1995 Unruh\cite{unruh81} showed (numerically and for $\xi=-1$) 
that they are not sensitive when $T_{ac} \ll c_s/ d_c$ 
and when the initial state is vacuum for $\Omega \gg T_{ac}$. 
These conclusions are now understood 
analytically, see \cite{tedriver} for a review.
In spite of this absence of asymptotic modification,
when propagating backwards in time a mode (or better a wave packet), 
one finds that its near horizon propagation is dramatically modified
when its blue-shifted frequency reaches the critical frequency 
$\Omega_c = c_s /d_c$, see \cite{bmps} for schematic representations
of the modified trajectories. 

Being insensitive to high energy physics, 
the asymptotic Hawking radiation is truly a low
energy $O(T)$ phenomenon.
This reinforces our confidence in the low energy predictions
of the semi-classical treatment but does not take us any further
concerning its dubious high energy predictions.

\newpage

\subsection*{Inflationary models}

In inflationary scenarios, the primordial density fluctuations 
find their origin in the non-adiabatic evolution of the 
non-homogeneous modes of the inflaton which were in their vacuum
state at the onset of inflation.
Given the exponentially growing redshift during inflation,
their initial frequency $\Omega$ was proportional to $e^{H \Delta t}$,
where $H$ is the Hubble constant and
$H \Delta t$ is the number of e-folds during inflation.
If the latter is taken to be larger than $\sim 70$ (which is the minimal 
number for inflation to work when the adiabatic era starts just below
GUT scale, near $10^{15}\,$Gev) there is an appeal to Planckian physics.
Therefore the doubts concerning Hawking radiation apply here as 
well\cite{tedriver}.
Thus the appeal to modified dispersion relations can also be 
called upon to see if the scale-invariance of the primordial fluctuations
spectrum is affected by modifications of Planck 
scale physics\cite{cosm}. 

As in black hole physics, the properties of the spectrum are robust.
The two conditions guaranteeing this robustness are
the adiabaticity of the evolution from Planck scale to that of $H$, 
and the regularity of the initial configurations.
The adiabaticity essentially follows from scale separation, 
$M_{Planck} \gg H$, whereas the second condition
is severely constrained by self-consistency in the following
sense\cite{cosm2}. The very existence of inflation implies 
that the dominant 
term in the Friedman equation is the inflaton potential energy.
Thus the contribution of the other modes must be negligible.
When $H \ll M_{Planck}$
and when inflation lasts more than $70$ e-folds,
this condition requires that the initial state of the modes 
observed today be vacuum.

\subsection*{Experiments ?}

Can we conceive and realize experiments in which Hawking
quasi-particles could be measured ? 
We used the term ``quasi-particles'' rather than phonons to
emphasize that in condensed matter there is a large variety 
of collective excitations which can propagate in effective metrics.
At present it is not clear which material is the most suitable
for our purpose.
Another challenge is to
build a regular horizon enclosing a trapped region. 
But probably the main difficulty is to get a sufficiently high
temperature. Indeed it does not seem 
easy to get $T_{ac} > 10^{-6}$K. 

To overcome the detection of these soft quanta,
one should perhaps consider a pair of acoustic holes.
A natural feature of acoustic holes is to come in pairs: 
it suffices to reduce the velocity flow $v_B$ so as to
re-cross the sound velocity. Moreover this second crossing 
leads to a white hole. And white holes are unstable. 
This instability would show up differently
according the sign of $\xi$. For positive  $\xi$, one would obtain 
a self-amplifying flux of soft phonons, a laser effect, 
see \cite{bhlaser}. On the other hand,
 for $\xi$ negative, the ``partners'' of Hawking quanta 
would leak out of the white horizon with a frequency close to the cutoff 
$\Omega_c = c_s/ d_c$ and would thus be easier to detect. 

\subsection*{Orienting research on Quantum Gravity ?}

Even though this is clearly a delicate enterprise and one 
which might turn out to be misleading, it
has been one of the major reasons to study acoustic holes. 
Let us take some risk and present our opinion as it stands today. 

Besides the insensitivity of asymptotic properties, 
we have also learned that for {\it all} dispersion relations
but the scale invariant one $d_c = 0$, there is no appeal to 
unbounded frequencies.
[When the gradient $dv_B/dx$ vanishes 
for $x-x_{horizon} \gg d_c$, the frequency of a backwards in time 
propagated wave packet stays of the order of $\Omega_c$ 
for both $\xi= \pm 1$, without any further blueshift. One learns from this 
that the stationarity of Hawking radiation now 
results, as in the Golden Rule, by a steady increase 
of the density of modes at some fixed frequency. 
This is not what is found when there is no dispersion, 
see Eq. (3.34) in \cite{GO}.]
This suggests that the quantum 
gravitational interactions neglected in the semi-classical treatment
might introduce a {\bf new} scale $d_{QG}$ which could act on light 
propagation near the event horizon as $d_c$ did on phonons. 
In order words, instead of questioning  the robustness of asymptotic 
properties, we are lead to question {\bf is $\Omega=k$ robust} ? 

I would claim probably not. Because $\Omega=k$ follows from Eq. (1) 
which results from a linearisation procedure: when starting from
a quantum formulation of black hole evaporation, Eq. (1)
is obtained by discarding all fluctuations of $g_{\mu \nu}$ 
around the mean metric, 
a solution of the semi-classical Einstein equations.
A first order analysis shows that the effects of 
metric fluctuations on
configurations giving rise to Hawking quanta 
grow like $1/(r-2M)$.
Therefore a taming mechanism is bound to occur. 
This has been substantiated in \cite{hawkmf}
where it is shown that vacuum energy fluctuations
engender metric fluctuations whose random properties 
induce a damping similar to that obtained
by putting $\xi = - i$ in Eq. (4).
That near horizon quantum gravitational interactions 
should induce dissipative effects was anticipated in \cite{GO}.

\subsection*{Acknowledgements}

I wish to thank the organizers of the {\it Journ\'ees Relativistes} 
in Dublin,
as well as those of the  {\it ESF-COSLAB} meeting in London for 
the invitation.  
I also wish to thank Ted Jacobson for many discussions. 
This work has been
supported by the NATO Grant CLG.976417.


\end{document}